\documentclass[final,3p,times]{elsarticle}

\usepackage{graphicx}
\usepackage{multirow}
\usepackage{natbib}
\usepackage{hyperref}
\usepackage{subfig}
\usepackage{xcolor}
\usepackage{amssymb}

\usepackage{lineno}
\usepackage{amssymb}

\journal{Nuclear Physics A}

\begin{document}

\begin{frontmatter}


\title{Role of neutron transfer in the reaction mechanism of \\ $^9$Be+$^{169}$Tm, $^{181}$Ta, $^{187}$Re and $^{197}$Au systems}
\author[a]{Prasanna ~M.} 
\author[b,c]{V.~V.~Parkar} \ead{Corresponding author: vparkar@barc.gov.in} 
\author[b,c]{V.~Jha} 
\author[d]{A.~Parmar} 
\author[a]{Bhushan.~A.~Kanagalekar} 
\author[a]{B.~G.~Hegde} 
\address[a]{Department of Physics, Rani Channamma University, Belagavi - 591156, India}
\address[b]{Nuclear Physics Division, Bhabha Atomic Research Centre, Mumbai - 400085, India}
\address[c]{Homi Bhabha National Institute, Anushaktinagar, Mumbai - 400094, India}
\address[d]{Department of Physics, Faculty of Science, The M. S. University of Baroda, Vadodara - 390002, India}

\begin{abstract}
The contribution of one neutron stripping cross section to the total reaction cross section has been studied for $^9$Be projectile incident on $^{169}$Tm, $^{181}$Ta, $^{187}$Re and $^{197}$Au targets around Coulomb barrier energy. The measured one neutron stripping cross sections for these systems have been compared with the coupled channel calculations. The recently developed global set of optical model potential parameters for $^9$Be projectile has been used in the present calculations. The cumulative of measured complete fusion (CF), incomplete fusion (ICF), one neutron stripping and calculated non-capture breakup (NCBU) cross sections is found to explain almost the reaction cross sections for all the targets. A very small contribution from target inelastic states and elastic breakup may contribute to the remaining part. The percentage fraction of cross section for CF, ICF, one neutron stripping, and NCBU over reaction cross section show the dominance of neutron transfer and NCBU at below barrier energies while CF and ICF processes have the major contribution at above barrier energies.
\end{abstract}

\begin{keyword}
Weakly bound nuclei, neutron transfer, reaction mechanism 

\end{keyword}

\end{frontmatter}


\section{\label{sec:Intro} Introduction}
Elastic scattering, fusion, transfer and breakup reactions and their relative importance on reaction  mechanism involving weakly bound projectiles is a  topic of interest for last few decades \cite{Jha20, Cant15, Kola16, Keel07, Keel09}. In this context, several experimental studies have been performed  over  the  years  utilizing  projectiles  of  both  stable  and unstable  weakly bound nuclei. The systematic of inclusive $\alpha$, reaction and total fusion cross sections with various  projectiles  have  been  highlighted  in these studies \cite{Jha20}. However, the contribution of  neutron transfer cross sections in reactions with these projectiles has not received  similar attention. The $^9$Be projectile, having a cluster structure  of $\alpha$ + $\alpha$ +  n and with  a  breakup threshold of 1.67 MeV, presents a case where the neutron  may play a major role in the reaction dynamics. In the case of $^9$Be reaction on some target, neutron stripping leads to $^8$Be, which disintegrates subsequently into two $\alpha$ particles, hence this process is expected to be a major contributor in the reaction cross section. Moreover, this process is also expected to contribute significantly to the $\alpha$ production apart from  $\alpha$ ICF channel in $^9$Be induced reactions. 

One neutron stripping reactions represent a powerful method for extracting the spectroscopic information
relating to single particle states. The most commonly used stripping reaction is (d,p) reaction, which is also the most well understood reaction from the theoretical point of view, owing to the simplicity of the light particles involved. One neutron stripping reactions induced by heavy ions such as $^7$Li, $^{11}$B, $^{12}$C, $^{14}$N, and $^{16}$O \cite{Sch73, Kundu17, Ford74, Pieper78, Santra01, Jha04} involve the transfer of a neutron that is fairly tightly bound in the projectile. Thus, the ground-state reaction Q values are generally negative and transitions to high-spin states at low excitation energies are favoured. Contrary to this, $^9$Be projectile, owing to a small neutron binding energy (S$_{n}$=1.67 MeV) and the resulting large positive Q values, populates low-spin states up to several MeV in excitation with considerable strength \cite{Jones2013}. Therefore, ($^9$Be, $^8$Be) reaction has been utilised to find out the spectroscopic factors for discrete states on several targets \cite{Stahel77}. Recently, One neutron transfer reactions have been used on $^9$Be target by inverse kinematics reactions at Holifield Radioactive Ion Beam Facility (HRIBF) \cite{Beene2011} at Oak Ridge National Laboratory, USA. Single neutron states in $^{131}$Sn, $^{133}$Sn and $^{135}$Te were populated in ($^9$Be, $^8$Be) reaction in inverse kinematics and studied using particle-$\gamma$ coincidence spectroscopy \cite{Jones22, Allmond14, Allmond12}. 

 In the recent work for $^9$Be+$^{169}$Tm, $^{181}$Ta and $^{187}$Re \cite{Fang16}, and $^{197}$Au \cite{Kaushik21} systems, one neutron stripping cross sections have been measured over a wide energy range around barrier. In addition, CF and ICF cross sections have also been reported \cite{Fang15, Zhang14, Kaushik20}. Thus a comprehensive data set exists that can be utilized for probing the reaction mechanisms of processes involved in systems with $^9$Be projectile. A  set of potential parameters for the description of elastic scattering angular distribution have been developed by Xu \textit{et al.} \cite{9Be_Xu}, which gives a nice description of elastic scattering data of $^9$Be projectile on a variety of target systems over a wide range of energies. These set of potential parameters can be utilized to get very reliable predictions of the reaction cross sections. Together with the measurement as described above, the relative importance of all the dominant processes can be estimated that can provide useful insights for reaction mechanism of systems involving weakly bound projectiles. In the present work, the coupled channel calculations for estimation of the neutron transfer and NCBU cross sections for these systems have been performed. The relative contribution of CF, ICF, neutron transfer and NCBU with respect to the reaction cross sections have also been deduced. The paper is organised as follows. Calculation details are given in Sec. II. The results are described in Sec. III, while discussion is given in Sec. IV. Summary is given in Sec. V. 

\section{\label{sec:Caln} Calculation Details}
Coupled reaction channel (CRC) calculations and Continuum Discretised Coupled Channel (CDCC) calculations have been performed using the code FRESCO \cite{Thom88} (version FRES 2.9) to estimate the one neutron transfer and NCBU cross sections respectively. Next, we discuss each method in brief.

\subsection{CRC calculations}
CRC calculations considering the one neutron transfer to various excited states of the residual nucleus have been performed. Optical model potential parameters required  for  entrance and exit channels were taken from recently developed global optical model potential for $^9$Be projectile \cite{9Be_Xu}. This potential was developed by systematically studying the experimental data of elastic scattering angular distributions and reaction cross sections for large number of targets in the mass range A=24-209. Apart from the optical model parameters, binding potentials of the fragment and core for the projectile and target partitions were used. The depths were adjusted to obtain the required binding energies of the particle-core composite system. These potential parameters employed in the calculations are also tabulated in Table\ \ref{tab1}. The single particle states of the residual nuclei taken from Ref.\ \cite{nndc} and listed in Table\ \ref{tab2} with spectroscopic factors (C$^2$S) of all the states as 1.0 were also considered. In the CRC calculations, the finite range form factors in the post form were used, including the full complex remnant terms and non orthogonality corrections.     

\subsection{CDCC and CDCC+CRC calculations}
The CDCC and CDCC+CRC calculations have been carried out to study the effect of breakup and transfer couplings as well as estimating the NCBU and transfer cross sections simultaneously. The CDCC calculations were performed considering $^9$Be as $^8$Be+n cluster structure. A two body $^8$Be+n cluster structure has been shown to describe the elastic scattering of $^9$Be on various target systems \cite{Pandit11, Parkar13}. The calculation details are given in our earlier works \cite{Pandit11, Parkar13, Jha14, Pals14b} and a brief summary is given here. 

The ground state wave function of $^9$Be is generated using a potential with Woods-Saxon volume type and a spin-orbit component taken from Refs.\ \cite{Lang77, Balzer77}. The 1/2$^+$ and 5/2$^+$ resonance states are generated by using the same potential parameters as that of the ground state except for the potential depth, which is varied to obtain the resonances with correct energies. The non-resonant continuum states are then generated with the same potential as that used for the resonance states. A depth of 44.9 MeV was found to reproduce the binding energy of ground state while 70.3 MeV was used for the resonant and non-resonant states. A relative angular momentum value of up to $\ell$ = 4 for the neutron core relative motion is taken for the calculations. The continuum up to an energy E$_{max}$ = 7 MeV above the $^8$Be + n breakup threshold was discretized into momentum bins of width $\Delta$k=0.1 fm$^{-1}$. The continuum discretization considering $\ell$ = 4 and E$_{max}$ = 7 MeV have been found to provide the converged cross sections. The cluster folded potentials required in the CDCC calculations for constructing the $^9$Be+target interaction potential are obtained using the two-body core-target and valence-target potentials V$_{^8Be-T}$ and V$_{n-T}$ respectively. For the V$_{n-T}$, the potentials are taken from Ref.\ \cite{Morillon07}, while for V$_{^8Be-T}$, the potential of $^9$Be is used \cite{9Be_Xu} for all the systems. To test cluster folded potential parameters, the elastic scattering data available for $^9$Be+$^{197}$Au system \cite{Gollan20} was utilised. The depth of the real part of both the n-T and $^8$Be-T was needed to be normalised by a factor of 0.8 along with the increase in radius of the real part of $^8$Be-T by 0.1 fm to fit the elastic scattering data at all the energies. In addition to the breakup couplings, the effect of neutron transfer couplings have also been investigated through a combined CDCC+CRC approach as explained in Refs.\ \cite{Pandit11, Parkar13}, where the CDCC wave functions are used for the transfer calculations. From these calculation, NCBU and neutron transfer cross sections have been estimated simultaneously. 

\begin{table*}
\caption{Potential parameters used in CRC calculations for $^{9}$Be+$^{169}$Tm, $^{181}$Ta, $^{187}$Re and $^{197}$Au systems. The radius parameter in the potentials are derived from R$_{i}$ = r$_{i}$ A$^{1/3}$, where i = R, V, S, SO, C and A is the target mass number.}
\begin{tabular}{cccccccccccccc}
\hline\hline\
System & V$_{R}$ & r$_{R}$  & a$_{R}$  & W$_{V}$  & r$_{V}$  & a$_{V}$  & W$_{S}$  & r$_{S}$  & a$_{S}$ & V$_{SO}$  & r$_{SO}$  & a$_{SO}$ & r$_{C}$\\
&(MeV)&(fm)&(fm)&(MeV)&(fm)&(fm)&(MeV)&(fm)&(fm)&(MeV)&(fm)&(fm)&(fm) \\
\hline\
$^{9}$Be+$^{169}$Tm \cite{9Be_Xu}$^a$ & 257.35  & 1.35 & 0.73 & 16.69 & 1.64 & 0.60 & 46.15 & 1.20 & 0.84& - & - & - & 1.56 \\
n+$^{169}$Tm \cite{Kaushik21} $^b$ & 50.00 & 1.23 & 0.65 & - & - & - & - & - & - & 6.00 & 1.23 & 0.65 & 1.25\\
n+$^{8}$Be \cite{Lang77} & 50.00 & 1.15 & 0.57 & - & - & - & - & - & - & 5.50 & 1.15 & 0.57 & 1.25\\
\hline \hline
\end{tabular}
\footnotesize{$^a$Same potential parameters are used for other systems, $^{9}$Be+$^{181}$Ta, $^{187}$Re and $^{197}$Au systems, except the radius of the volume real part, It changes according to A of target} \\
\footnotesize{$^b$Same binding potentials for n+$^{181}$Ta, $^{187}$Re and $^{197}$Au systems}
\label{tab1}
\end{table*}

\begin{table}
\begin{center}
\caption{\label{specfac} Energy levels of residual nuclei and corresponding spin-parity values \cite{nndc} used in the CRC calculations. For $^{198}$Au residue, the same states as listed in Ref.\ \cite{Kaushik20} are used.}\ \\
\begin{tabular}{|cc|cc|cc|} 
\hline \multicolumn{2} {|c|} {$^{170}$Tm}&
\multicolumn{2} {|c|} {$^{182}$Ta}&
\multicolumn{2} {|c|} {$^{188}$Re} \\ \hline E& J$^{\pi}$& E& J$^{\pi}$& E& J$^{\pi}$
\\(MeV)&&(MeV)&&(MeV)&\\ \hline
0.000 & 1$^-$& 0.000 & 3$^-$& 0.000 & 1$^-$\\
0.039 & 2$^-$& 0.098 & 4$^-$& 0.062 & 2$^-$\\
0.115 & 3$^-$& 0.115 & 4$^-$& 0.154 & 3$^-$\\
0.150 & 0$^-$& 0.174 & 5$^-$& 0.170 & 3$^-$\\
0.183 & 4$^-$& 0.240 & 5$^-$& 0.182 & 4$^-$\\
0.204 & 2$^-$& 0.250 & 3$^+$& 0.205 & 2$^-$\\
0.220 & 2$^-$& 0.270 & 5$^+$& 0.231 & 3$^+$\\
0.237 & 1$^-$& 0.317 & 6$^-$& 0.260 & 2$^-$\\
0.270 & 3$^-$& 0.335 & 7$^+$& 0.287 & 4$^-$\\
0.320 & 5$^-$& 0.361 & 3$^-$& 0.300 & 2$^+$\\
0.350 & 3$^-$& 0.390 & 6$^+$& 0.321 & 4$^-$\\
0.360 & 4$^-$& 0.396 & 6$^-$& 0.335 & 5$^-$\\
0.381 & 4$^-$& 0.403 & 2$^+$& 0.343 & 4$^+$\\
0.440 & 5$^+$& 0.411 & 6$^+$& 0.360 & 2$^-$\\
0.450 & 3$^-$& 0.444 & 1$^-$& 0.372 & 3$^-$\\
0.467 & 5$^-$& 0.480 & 4$^-$& 0.440 & 3$^+$\\
0.540 & 4$^-$& 0.490 & 6$^-$& 0.462 & 4$^-$\\
0.550 & 5$^-$& 0.506 & 5$^+$& 0.483 & 5$^+$\\
0.608 & 3$^+$& 0.550 & 3$^-$& 0.500 & 3$^+$\\
0.650 & 1$^-$& 0.566 & 3$^-$& 0.512 & 4$^+$\\
0.693 & 2$^-$& 0.572 & 4$^+$& 0.521 & 4$^-$\\
0.715 & 3$^-$& 0.579 & 7$^+$& 0.555 & 1$^-$\\
0.750 & 3$^-$& 0.581 & 7$^-$& 0.576 & 3$^+$\\
0.774 & 3$^-$& 0.630 & 5$^-$& 0.582 & 1$^-$\\
0.822 & 2$^+$& 0.651 & 4$^-$& 0.606 & 2$^-$\\
0.854 & 2$^-$& 0.673 & 6$^+$& 0.628 & 2$^-$\\
0.908 & 2$^-$& 0.702 & 3$^-$& 0.641 & 5$^-$\\
0.921 & 5$^-$& 0.724 & 3$^+$& 0.655 & 5$^+$\\
0.960 & 3$^-$& 0.750 & 3$^+$& 0.674 & 3$^-$\\
0.980 & 3$^-$& 0.780 & 7$^-$& 0.696 & 5$^+$\\
1.014 & 4$^-$& 0.805 & 6$^-$& 0.703 & 4$^+$\\
1.140 & 2$^-$& 0.835 & 3$^-$& 0.710 & 5$^-$\\
1.291 & 4$^-$& 0.860 & 4$^-$& 0.723 & 2$^-$\\
1.395 & 2$^-$& 0.910 & 5$^-$& 0.740 & 3$^-$\\
1.448 & 3$^-$& 0.960 & 5$^-$& 0.781 & 5$^-$\\
1.537 & 1$^-$& 1.101 & 5$^-$& 0.800 & 5$^+$\\
      &      & 1.114 & 5$^-$& 0.832 & 2$^-$\\
      &      & 1.116 & 7$^-$& 1.012 & 4$^-$\\
      &      & 1.220 & 2$^-$& 1.086 & 4$^-$\\
      &      & 1.360 & 5$^-$& 1.147 & 2$^-$\\
      &      & 1.445 & 4$^-$& 1.190 & 4$^-$\\
      &      & 1.571 & 2$^-$& 1.232 & 5$^-$\\
      &      & 1.680 & 4$^-$&       &      \\
      &      & 1.747 & 4$^-$&       &      \\
\hline
\end{tabular}
\label{tab2}
\end{center}
\end{table}

\section{\label{sec:Result} Results}
\subsection{\label{sec:1nstrip}One neutron transfer}
\begin{figure}
\includegraphics[width=85 mm,trim=6.7cm 2.6cm 5.6cm 1.0cm,clip=true]{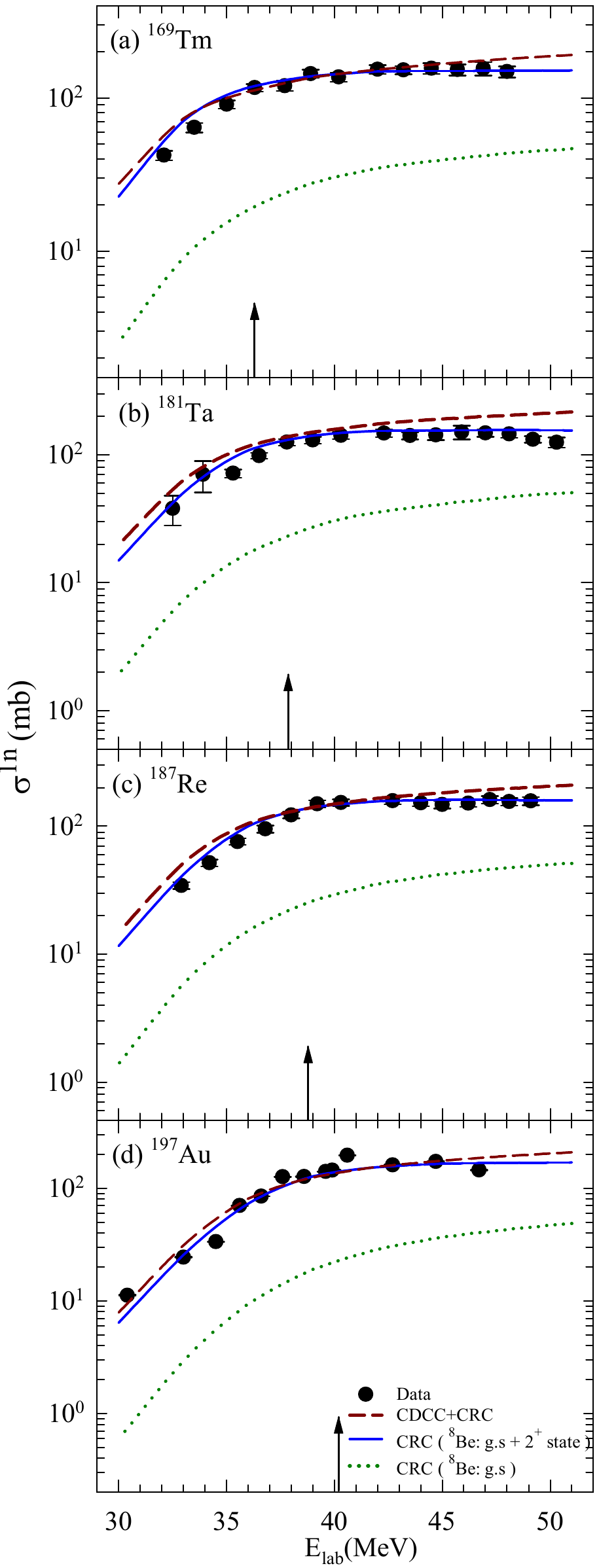}
\caption{\label{1nstrip}The comparison of measured one neutron stripping cross sections for $^9$Be+ (a)$^{169}$Tm, (b)$^{181}$Ta, (c)$^{187}$Re and (d)$^{197}$Au systems with the CRC and CDCC+CRC calculations (see text for details). The arrow indicates the Coulomb barrier position.}
\end{figure}

The one neutron stripping data available \cite{Fang16} for $^9$Be+$^{169}$Tm, $^{181}$Ta, $^{187}$Re and $^{197}$Au systems is compared with the present CRC only and CDCC plus CRC combined calculations in Fig.\ \ref{1nstrip}(a-d). CRC calculations show a good agreement with the data at all the energies while CDCC+CRC calculations slightly overpredict the data at few energies above the barrier. The cross sections corresponds to sum of all the channels of residual nucleus in which the $^8$Be ejectile is found in the ground state and ground plus 2$^+$ excited state respectively. Similar CRC only calculations and agreement with the data have already been reported for $^9$Be+$^{197}$Au \cite{Kaushik21} and $^9$Be+$^{159}$Tb \cite{Kaushik21a} systems. Major contribution of 2$^+$ resonance state (E = 3.03 MeV) of $^8$Be to the transfer cross section compared to ground state is seen as also pointed out in Refs.\ \cite{Kaushik21, Kaushik21a}.    

\subsection{\label{sec:Reac} Reaction mechanisms}
\begin{figure}
\includegraphics[width=85mm,trim=6.5cm 2.5cm 6.0cm 1.2cm,clip=true]{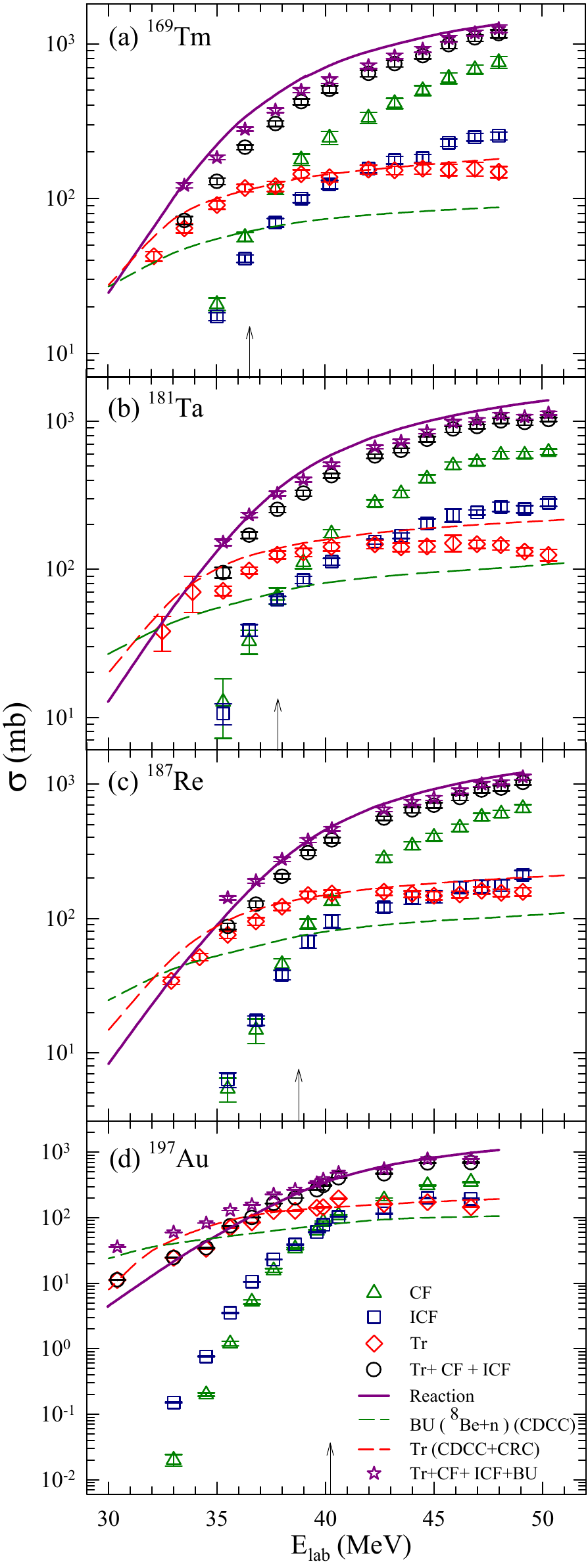}
\caption{\label{reaction}The measured CF, ICF, one neutron transfer, calculated NCBU and their sum is compared with reaction cross sections for $^9$Be+ (a)$^{169}$Tm, (b)$^{181}$Ta, (c)$^{187}$Re and (d)$^{197}$Au systems. The one neutron transfer cross section calculated in CDCC+CRC method is also shown in dashed line. The arrow indicates the Coulomb barrier position.}
\end{figure}
\begin{figure}
\includegraphics[width=85mm,trim=5.9cm 4.0cm 6.4cm 0.4cm,clip=true]{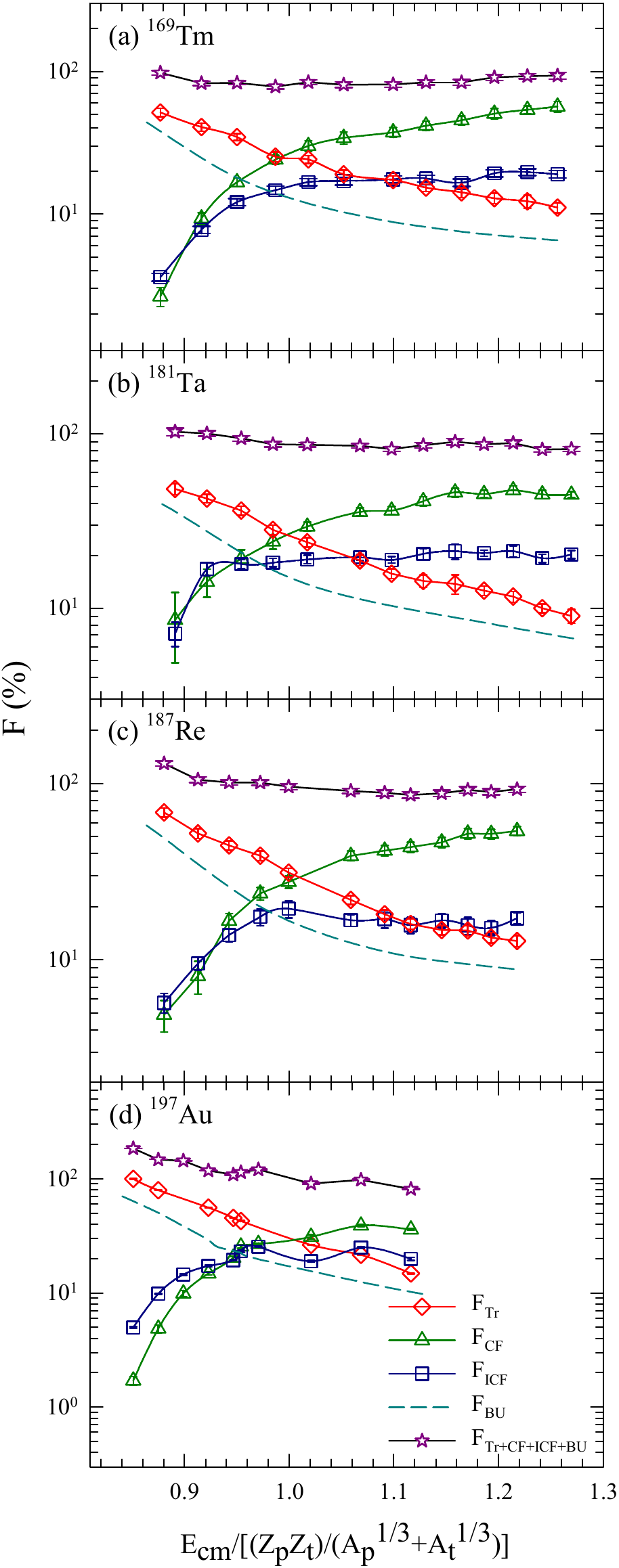}
\caption{\label{fraction}The relative fraction of measured CF, ICF and one neutron transfer data and calculated NCBU over reaction cross sections along with cumulative of all fractions for $^9$Be+ (a)$^{169}$Tm, (b)$^{181}$Ta, (c)$^{187}$Re and (d)$^{197}$Au systems. Lines for CF, ICF, 1n transfer and cumulative fractions are guide to an eye.}. 
\end{figure}
The CF and ICF cross sections were measured in $^9$Be+$^{169}$Tm, $^{181}$Ta, $^{187}$Re and $^{197}$Au systems \cite{Zhang14, Fang15, Kaushik20}. As the one neutron transfer cross sections were also measured in these systems at the same energies \cite{Fang16, Kaushik21}, we have taken the cumulative of measured CF, ICF, 1n transfer cross sections along with calculated NCBU cross sections and shown in Fig.\ \ref{reaction}(a-d). Measured individual CF, ICF and transfer cross sections and calculated NCBU cross sections are also shown in that figure. The NCBU and transfer cross section calculations are also shown as dashed dot dot and dashed lines respectively in Fig.\ \ref{reaction}(a-d). The reaction cross sections estimated with the global optical model potential \cite{9Be_Xu} that has been used in the present calculations is shown as a solid line. The reaction cross sections are slightly larger than the cumulative of fusion, neutron transfer and NCBU cross sections at above barrier energies for all the systems. It may have possible contributions from other channels such as elastic breakup and target inelastic states which were not measured/estimated.  

To understand the relative fraction of different reaction channels around the barrier, ratio of cross sections between measured CF, ICF and 1n transfer over reaction cross sections were plotted as a function of reduced energy (equivalent to E$_{c.m}$/V$_b$) in Fig.\ \ref{fraction}. We define the percentage fraction as,  $F_{CF/ICF/Tr}(\%)=\frac{\sigma_{CF/ICF/Tr}}{\sigma_{Reac}}* 100$. The fractions $F_{Tr}$, $F_{CF}$ and $F_{ICF}$ as a function of reduced energy are plotted in Fig.\ \ref{fraction}(a-d). Similarly calculated NCBU fraction is shown as $F_{BU}$. As can be seen, neutron transfer and NCBU are dominant at below barrier energies and these processes decrease with the increase in energy. The CF contribution is found to increase with beam energy and it is dominant at above barrier energies while the ICF contributes between 5-20 \% for all the targets. ICF cross sections are found to have larger contributions than neutron transfer and NCBU at above barrier energies. The cumulative fraction is also shown by solid line in Fig.\ \ref{fraction}, which suggests that the reaction mechanism is completely explained by cross sections due to these four processes except for a very small (10\%) difference that may be ascribed to  the inelastic excitations of the target which were not estimated in the present work. 

\section{\label{sec:Discussion} Discussion}
Systematic of neutron stripping cross sections with $^9$Be projectile on different targets was given in Ref.\ \cite{Kaushik21a}, which showed a target independence. It must be mentioned that similar feature is also observed for $^6$Li \cite{Parkar22} and $^7$Li projectiles \cite{Parkar21}. This implies that for neutron stripping reactions in these weakly bound stable projectiles, the detailed target structure may be inconsequential and major effect comes from the projectile structure. In fact, a dependence on the neutron separation energy for different projectile systems can easily be deduced from these observations. In the work of Ref.\ \cite{Kaushik21} on $^{197}$Au target with $^{6,7}$Li, $^9$Be and $^{10}$B projectile, neutron transfer cross section as a function of E$_{c.m.}$/V$_b$ was shown, which clearly indicates that the transfer cross section was dependent on neutron separation energy of the projectile, highest cross section with $^9$Be (S$_n$=1.67 MeV) and lowest with $^{10}$B (S$_n$=8.44 MeV) projectile. It will be interesting to further investigate this feature for a projectile system where the S$_n$ values are intermediate, \textit{viz.} $^{13}$C (S$_n$=4.95 MeV) projectile and also with unstable nuclei having lower S$_n$ values. Also, similar studies of neutron transfer with different targets will be useful. 

The present work further shows that neutron transfer is dominant mechanism and major contributor in reaction cross section in the case of $^{9}$Be projectile at energies below the Coulomb barrier. At these energies, this has more contribution as compared to CF and ICF processes. The latter processes gradually increase and dominate at higher energies with the cross over happening at $\approx$ $V_b$ which again is almost similar for all the target systems considered. This behaviour can be contrasted from that of $^{6,7}$Li projectiles, where the neutron transfer is smaller than ICF even at below barrier energies \cite{VVP18, VVP18b, Parkar21, Pals14, Ara13, Shrivastava09}. Thus the present study has clearly brought out the differences among the different stable weakly bound nuclei. These differences may be related to the cluster structures of $^9$Be ($^8$Be+n) compared $^{6}$Li ($\alpha$+d) and $^{7}$Li ($\alpha$+t) nuclei. Hence, the 1n transfer in $^9$Be could find an environment more favorable as compared to the $^6$Li and $^7$Li cases. 

\section{\label{sec:Sum} Summary}
In summary, CRC calculations have been performed using the reliable global optical model potential parameters of $^9$Be projectile for investigating the role of neutron transfer around barrier energies in $^9$Be+$^{169}$Tm, $^{181}$Ta, $^{187}$Re and $^{197}$Au systems. In addition, CDCC and CDCC+CRC calculations have been carried out to estimate the NCBU and neutron transfer cross sections simultaneously. Calculated one neutron stripping cross sections were found to explain the measured data for all the systems. The estimated reaction cross sections using the global optical model potential parameters for $^9$Be were found to exhaust almost the sum of measured CF, ICF, one neutron stripping and calculated NCBU cross sections. Relative fraction of CF, ICF, transfer, and NCBU revealed that neutron  transfer and NCBU are dominant reaction  channels at  below  barrier and CF dominates at above barrier energies, while ICF contributes between 5-20\% for all the targets. This feature is in contrast with other projectile systems such as $^{6,7}$Li nuclei, where ICF and CF provide the primary contribution to the reaction cross section even at below barrier energies.

\section{Acknowledgments}
P.M. acknowledges the financial support of Board of Research in Nuclear Science (BRNS), India (Sanction No: 58/14/04/2019-BRNS/10254) in carrying out these investigations. We are thankful to Mr. Anilkumar Naik from Computer Center and Communication Facility, TIFR, Mumbai for technical support to run our memory intensive calculation at the High-Performance Computing (HPC) cluster facility.


\end{document}